# Experimental Detection of the Gravitomagnetic London Moment

Martin Tajmar[1,*], Florin Plesescu[1], Klaus Marhold[1] & Clovis J. de Matos[2]

[1]*Space Propulsion, ARC Seibersdorf research GmbH, A-2444 Seibersdorf, Austria*

[2]*ESA-HQ, European Space Agency, 8-10 rue Mario Nikis, 75015 Paris, France*

It is well known that a rotating superconductor produces a magnetic field proportional to its angular velocity. The authors conjectured earlier, that in addition to this so-called London moment, also a large gravitomagnetic field should appear to explain an apparent mass increase of Niobium Cooper-pairs. This phenomenon was indeed observed and induced acceleration fields outside the superconductor in the order of about 100 µg were found. The field appears to be directly proportional to the applied angular acceleration of the superconductor following our theoretical motivations. If confirmed, a gravitomagnetic field of measurable magnitude was produced for the first time in a laboratory environment. These results may open up a new experimental window on testing general relativity and its consequences using coherent matter.

* Corresponding Author, Email: martin.tajmar@arcs.ac.at

## 1. INTRODUCTION

A superconductor exhibits two main magnetic properties. One is to expel any magnetic field from its interior (Meissner–Ochsenfeld effect) and the second one is to produce a magnetic field (London moment) when set into rotation. Derived from the quantization of the canonical momentum, the London moment is expressed by:

$$\vec{B} = -\frac{2m^*}{e^*}\vec{\omega}\ , \tag{1}$$

where $m^*$ and $e^*$ are the Cooper-pair mass and charge respectively. By precisely measuring the magnetic field and the angular velocity of the superconductor, one can calculate the mass of the Cooper-pairs. This has been done for both classical and high-$T_c$ superconductors[1-4]. In the experiment with the highest precision to date, Tate et al[5,6] reported a disagreement between the theoretically predicted Cooper-pair mass in Niobium of $m^*/2m_e = 0.999992$ and her experimental value of 1.000084(21), where $m_e$ is the electron mass. This anomaly was actively discussed in the literature without any apparent solution[7-10].

Two of the authors proposed that in addition to the classical London moment, a gravitational analogue exists. This so-called gravitomagnetic London moment could explain the Tate measurements[11-13]. One can then replace Equ. (1) by

$$\vec{B} = -\frac{2m^*}{e^*}\cdot\vec{\omega} - \frac{m^*}{e^*}\cdot\vec{B}_g\ , \tag{2}$$

where $B_g$ is the gravitomagnetic field. In order to explain the apparent mass increase from Tate's measurement, a gravitomagnetic field (gravitomagnetic London Moment) of

$$\vec{B}_g = 2\vec{\omega}\left(\frac{\Delta m}{m^*}\right) = 1.84 \times 10^{-4}\,\vec{\omega} \tag{3}$$

is required, where $\Delta m$ is the difference between the measured Cooper-pair mass and its theoretical prediction. This value is many orders of magnitude higher than any classical gravitomagnetic field (e.g. $B_{g,Earth} \cong 10^{-14}$ rad.s$^{-1}$)[14].

In subsequent publications, the authors showed that the London moment can be interpreted as a consequence of a photon gaining mass inside the superconductor via the Higgs mechanism due to the breaking of gauge symmetry. Consequently, it was shown that the gravitomagnetic London moment has its origin in a non-zero graviton mass inside a superconductor[13]. In a follow-up paper[15], it is shown that the graviton mass calculated from the gravitomagnetic Larmor theorem gives rise to a gravitomagnetic London moment which is within about an order of magnitude to the one predicted from Tate's measurements. We then predict a gravitomagnetic London moment of

$$\vec{B}_g = 2\vec{\omega}\left(\frac{\rho_m^*}{\rho_m}\right), \tag{4}$$

where we can easily interpret the ratio of $\Delta m/m^*$ as the ratio between the Cooper-pair mass density and the bulk density of the material.

This gravitomagnetic London moment is equivalent to a frame-dragging or Lense-Thirring field and is of great interest since gravitational effects of measurable magnitudes could be produced in a laboratory environment.

Indeed, in the past many attempts were done to find Lense-Thirring fields at the laboratory scale (e.g. Ref. 16,17). However, all these approaches failed and only space missions[18] such as Gravity-Probe B currently allow measurements of this prediction of general relativity. In the present paper, we describe our apparatus and report the first measurements which show strong evidence that indeed a gravitomagnetic London moment, following our motivations to explain the missing Cooper-pair mass, exists. If confirmed, this opens up new means of investigating general relativity and its consequences in the quantum world using coherent matter.

## 2. MEASUREMENT CONCEPT

An experiment was designed to measure the gravitational field induced by a non-stationary gravitomagnetic field generated by an angularly accelerated superconducting ring. By applying an angular acceleration to the ring, following the gravitational induction law[19,20]

$$rot\,\vec{g} = -\frac{\partial \vec{B}_g}{\partial t} \quad , \tag{5}$$

a non-Newtonian gravitational field, as opposed to a Newtonian-like divergent field, is generated along the tangential direction (azimuthal plane). Using Equ. (4) we can express the gravitational field expected at radial distance *r* inside the superconductor ring as

$$g = -\dot{B}_g \frac{r}{2} \,.\, \hat{\varphi} = -\frac{\rho_m^*}{\rho_m} r\dot{\omega} .\, \hat{\varphi} = -\frac{2.624 \times 10^{-18}}{\lambda_L^2 \rho_m} r\dot{\omega} \,.\, \hat{\varphi} \,, \tag{6}$$

with $\lambda_L$ as the London penetration depth, $\hat{\varphi}$ the azimuthal unity vector and $g$ is measured in units of the Earth's standard acceleration. The gravitational field, following the induction law in Equ. (5), should point in the opposite direction of the applied tangential angular acceleration.

## 3. EXPERIMENTAL SETUP

Following Tate's experiment, we tested a Niobium ring (dimensions 150x138x15 mm), as well as a lead sample (classical Type-I superconductor in contrast to the Niobium Type-II with same dimensions), an YBCO ring (160x130x15 mm, manufactured by ATZ) and a BSCCO (2212) ring (150x138x15 mm, manufactured by NEXANS).

As shown in **Fig. 1**, the superconductor is mounted in the middle of a large cryostat. Two types of motors (Torquemaster BNR3034 brushless server motor and a Düsterloh PMW 400 Z24 ML/MR pneumatic air motor) can drive the superconductor ring up to 6500 RPM. The speed was measured by an encoder mounted in the electric motor and an additional optical encoder, which was necessary for the air motor. By comparison between commanded and measured speeds, the maximum deviation was < 5% at speeds up to 1000 RPM and < 1% at full speed. Around the superconductor, a coil was mounted to test the influence of field trapping on our results. The accelerometers to measure the gravitational field induced by the rotating superconductor are mounted inside an evacuated chamber made out of stainless steel which acts as a Faraday cage and is directly connected by three shafts to a large structure made out of steel that is fixed to the floor and roof of the building. The sensors inside this chamber are thermally isolated from the cryogenic environment due to the evacuation of the sensor chamber and additional MLI isolation wrapped around the sensors. Only flexible tubes along the shafts (necessary to seal the cryostat) and electric wires from the sensor chamber to the upper flange establish a mechanical link between the sensor chamber and the cryostat. This system enables a very good mechanical de-coupling of the cryostat with the rotating superconductor and the accelerometers even during high rotational speeds. In order to damp the oscillations due to the rotation of the superconductor, the whole cryostat is mounted in a box filled with 1500 kg of sand (see **Fig. 2**).

Within the sensor vacuum chamber, three AppliedMEMS 1500S accelerometers (radial, tangential and vertical direction) were mounted on the three measurement positions (9 in total) as shown in **Fig. 3**. The in-ring position should give the highest field strength according to the gravitational induction law. We also included an above-ring position in order to study how the field propagates. The reference position monitors the mechanical movements and facility artifacts, if the gravitomagnetic London moment propagates similar to electromagnetic fields, then the signal from the rotating superconductor should be less than 2 orders of magnitude at the reference location compared to the in-ring position.  The accelerometers used provide one of the lowest noise level available (300 $ng_{rms}/\sqrt{Hz}$) at a compact size so that they can be mounted inside the superconductor; moreover, they are insensitive to magnetic fields in order to minimize any interference with the electric motor. The sensors were delivered with calibration certificates and the calibration value from each individual sensor was taken into account during the measurements. The calibration was also checked for each sensor by re-orientating the sensor chamber to measure the Earth’s gravitational field with each sensor. On each level (in-ring, above-ring and reference position), PT-100 temperature resistors and Kapton heating foils provided temperature control to the accelerometers and maintained them at 25°C throughout the measurements. In addition to the accelerometers, high-resolution fluxgate magnetic field sensors (Stefan Mayer Instruments FL1-100) were mounted on the In-Ring and Reference positions. In order to obtain a reliable temperature measurement, a calibrated silicon diode (DT-670B-SD from Lakeshore) was installed directly below the superconductor. A miniature collector ring (MD6038 and MD6043 from LTN Precision Products) on top of the axis enabled the correct readout even during high speed rotation. The temperature stayed constant to within 0.1 K at speeds up to 6500 RPM.

The accelerometers/magnetometers were read out using Keithley 2182 nanovoltmeters with a measurement rate of 10 Hz. In order to measure all acceleration positions at the same time, each sensor was connected to its own nanovoltmeter. The accelerometer sensors are connected to an electric circuit that subtracts the reference signals from the in-ring and above-ring signals in order to filter out any mechanical noise vibrations, before being fed to the nanovoltmeters. All non-transition measurements were carried out in this differential configuration to filter out any mechanical movements of the sensor chamber (e.g. tilts), especially during the gas expansion phase when rotating in the liquid nitrogen and liquid helium environment.

## 4. EXPERIMENTAL RESULTS

### A. Initial Testing and Identification of External Sensor Influence

In order to reach the highest accuracy, the motor speed and acceleration were set to their limits. With the electric motor, a top speed of 6500 RPM accelerated and decelerated in 1 s could be achieved. The air motor had a lower top speed of about 4500 RPM but could accelerate and decelerate much faster by reversing the direction of the air flow.

At rest, the sensors using differential readout in vertical position had a noise level with a sigma of 5 µg and the radial and tangential sensors of about 15 µg. Applying an angular acceleration of 650 rad.s$^{-2}$ with the electric motor, no significant change was seen on the sensor output, verifying our mechanically decoupled setup. This gave us a coupling factor (induced gravitational field measured in units of *g* versus applied angular acceleration) noise level of about $\pm 2.10^{-8}$ s$^2$, valid for all rings (YBCO, BSCCO and Niobium) used during the tests. Comparing this noise level with our theoretical expectations, the anomaly derived out of Tate's measurement with a coupling factor of $-37\mathrm{x}10^{-8}$ s$^2$ should be clearly visible. However, this noise level is above our predictions from the Cooper-pair and bulk density ratio which leads to an expected coupling factor for Niobium of $-0.7\mathrm{x}10^{-8}$ s$^2$.

The following external sensor influences were found:

- As soon as the low-temperature bearings degraded or if the rotating axis got in resonance (usually at an angular velocity of about 400 rad.s$^{-1}$), the resulting acoustic noise triggered a negative offset on all sensors. After continued use of a damaged bearing, offsets in the order of milli-g were observed. However, as the noise level is independent of the sense of rotation, the offsets can be clearly identified by always performing two slope measurements after each other with alternating sense of rotation.

– During the liquid helium experiments, the high rotational speeds of the superconductor triggered excessive evaporation of the liquid helium in the cryostat. This resulted in an overpressure of about 0.5 bar, which induced noise peaks on the sensors due to the connection of the outer surface of the sensor chamber with the helium gas. The problem was reduced by maintaining a distance of at least 20 cm between the rotating superconductor and the liquid helium level in the cryostat.

– Using the coil, the influence of a strong magnetic field on the accelerometers was evaluated. The test was done with the BSCCO ring at 117 K (normal conductive) and 77 K (superconductive). An oscillating magnetic field with an amplitude of 20 mT was applied and the sensor responses were evaluated. The sensor offsets were found to be linearly proportional to the applied field, at 20 mT a maximum offset of 10 µg was measured. As the maximum magnetic field from the electric motor was measured to be about 50 µT, the maximum offset for the real runs should be therefore less than 0.025 µg, which is far below the measurement threshold of the sensors (about 1 µg). Therefore, influence of magnetic fields on the results can be neglected.

A summary of the uncertainties for each component used throughout the measurements is shown in **Table 1**.

## B. Testing with Liquid Nitrogen

Niobium, BSCCO and YBCO were tested at 77 K and 300 K with angular accelerations up to 650 rad.$s^{-2}$ (Niobium was tested as a reference sample since it is not superconductive at 77 K). All sensors show no difference within the noise level between both temperatures during the test runs. Especially within the acceleration time periods, no signal shifts, which would indicate the presence of a gravitomagnetic London moment, could be detected. All measurements were performed more than 20 times with no significant correlation factors plotting the measured gravitational field against angular speed, angular speed^2 (to test for centrifugal forces) or angular accelerations.

Therefore, we can conclude, that no detectable gravitational fields are present during testing with High-$T_c$ superconductors within our measurement accuracy. Although the resolution was low enough to test our Tate Cooper-pair mass anomaly hypothesis, this result was to be expected from Equ. (4) as the Cooper-pair densities in High-$T_c$ superconductors are much lower than the one in Niobium (as in the case of Tate). Therefore, the coupling factor noise level of $\pm 2.10^{-8}$ $s^2$ can be interpreted as an upper limit for any possible induced gravitational fields in this case.

The tests showed that the facility gives reliable results within the coupling factor noise level with no measurable influence from temperatures down to 77 K.

## C. Testing with Liquid Helium

### *1. YBCO*

As a reference case, the YBCO sample was also tested down to liquid helium temperatures in order to evaluate any additional side effects due to the change of coolant. No changes in the gravitational field measured by all sensors during the acceleration periods were seen within the sensor's resolution, leaving again a coupling factor noise level of $\pm 2.10^{-8}$ $s^2$. We conclude that the facility continues to give reliable low-noise measurements even during liquid helium operation.

### *2. Niobium*

Next, the Niobium superconductor ring was tested again at liquid helium temperatures. Contrary to our tests at room temperature and 77 K, in this case a clear change in the tangential acceleration could be measured by the sensors. **Fig. 4** shows how the in-ring tangential sensors (differential with respect to the reference sensors) react to the applied angular acceleration from an air motor – they react in the opposite direction as predicted from Tate's measurement and our hypothesis to correct it. As the rotating shaft has a resonance frequency of about 400 $rad.s^{-1}$, the sensor signals were damped by a factor of 5 for angular velocities above 350 $rad.s^{-1}$ to reduce the mechanical vibration noise. All acceleration peaks as shown were obtained when the angular velocity was below 350 $rad.s^{-1}$, which means that the measured peaks are not influenced by this damping at all. In a sense, the acceleration fields that were observed can be interpreted as a gravitational variant of Lenz's law. A similar behavior was also seen on the above-ring sensor. Note that at a temperature of 10 K, close above Niobium's critical temperature of 9.4 K, no effect in the gravitational field is measurable any more (the silicon diodes have a tolerance of ± 0.5 K). Only applying a tangential acceleration of about 1500 $rad.s^{-2}$ to the Niobium ring while it is superconductive results in a tangential acceleration measured outside the superconductor by the sensors (shielded from the superconductor in an evacuated massive Faraday cage) of around 100 µg in opposite direction. That is a clear indication of the gravitomagnetic London moment and its induced gravitational field following the gravitational induction law as we predicted. These values result in a coupling factor of about $-6.6 \times 10^{-8}$ rad.s$^{-2}$. Unfortunately, the peaks measured are still close to the noise of the sensors (signal-to-noise ratio is about 3.3) although the accelerometers feature the lowest noise level presently available. Since the peaks are "only" 100 µg, they have previously not been observed in measurements for the classical London moment. Nevertheless, they are 30 orders of magnitude higher than what general relativity predicts classically[13] and are therefore of great technological and scientific interest.

The acceleration fields also appear when performing a temperature transition over Niobium's critical temperature of 9.4 K while rotating as shown in **Fig. 5 (left)**. When the superconductor passes its critical point, a strong peak in the tangential acceleration measured by the in-ring and above-ring sensors is clearly visible. Note that no peak is visible any more at the reference tangential sensor location, which is a proof that the field is actually emitted from the superconductor itself. No peak is seen when the temperature falls below $T_c$ but the superconductor is not rotating any more. Also the direction of the peak is changing with the sense of rotation. With our measurement rate of 10 Hz, the peak of 277 µg at an angular velocity of about 550 rad.s$^{-1}$, we can estimate a lower limit (the temperature transition happened during a period of 0.1 s) for the coupling factor of about $5 \times 10^{-8}$ s$^{2}$, a value similar to what we got above from applying an angular acceleration while the Niobium ring was superconductive but with a different sign. The reason for the sign change is not clear yet, as we would have expected that the gravitational peak is following the sign of the angular speed according to the induction law in Equ. (6). Nevertheless, the fact that such a gravitational peak is measured when passing $T_c$ is a strong indication that our effect indeed happens only in the superconductive state.

Many measurements were conducted over a period from June to November 2005 to show the reproducibility of the results. **Fig. 6** summarizes nearly 200 peaks of in-ring and above-ring tangential accelerations measured by the sensors and angular accelerations applied to the superconductor as identified e.g. in **Fig. 4** with both electric and air motor. The correlation factors for the resulting slopes are both greater than 0.96, a clear indication that there is indeed a direct proportionality between applied acceleration and measured accelerations, as it should be according to the gravitational induction law. Hence, our concept of analyzing the data in terms of a coupling factor as defined above makes sense. Within our sensor resolution, the effect seemed stable for temperatures up to $T_c$ similar to the classical London moment. According to Equ. (4), we expected that the magnitude would change according to the Cooper-pair density, however, the scatter is still too high to definitely test such behavior. Moreover, the temperature measurement is done on the bottom of the superconductor. Together with the fact that during one profile the temperature changes in that point on the average about 1 K, and the tolerance of the silicon diode of 0.5 K, makes a clear identification of the Cooper-pair density slope difficult. Above $T_c$, no effect could be seen which rules out a coupling factor up to $\pm$ 2x$10^{-8}$ $s^2$ as also stated above.

It is important to note that the average coupling factor for the above-ring sensor over all measurements is 90% of the in-ring sensor, a value which would be expected from a field which is expanding similar to electromagnetism. This is a clear indication that the gravitomagnetic London moment is emitted from the superconductor. The coupling factor value of $-9.64\times10^{-8}$ $s^2$ is about a factor 13 higher than predicted from Equ. (4) and a factor of 3.8 smaller as derived from Tate's measurements. Therefore, despite this deviation, we can conclude that our hypothesis to correct Tate's derived Cooper-pair measurement with gravitomagnetic fields is justified.

### *3. Lead*

In order to investigate if the results above also appear in other low-$T_c$ superconductors, further measurements were conducted with lead. According to Equ. (4), we expected a gravitomagnetic London moment which was 84% the value from the one obtained with Niobium because its higher bulk density leads to a larger graviton mass. Indeed, the gravitomagnetic London moment was again observed on the in-ring and above-ring sensors only.

**Fig. 5 (right)** shows a temperature transition measurement. Here we see again a gravitational peak in the tangential in-ring and above-ring sensors when the superconductor passes its critical temperature at 7.2 K during rotation. Note that the direction of the peak is similar to the Niobium case but with a smaller value. The fact that the peak now appears at the critical temperature of lead is also a good indication that the effects we see are no artifacts from the liquid helium but correspond to the superconductor's properties.

**Table 2** summarizes our findings for Niobium and Lead and the High-$T_c$ superconductors. More than 50 measurements for the in-ring and above-ring tangential locations show that indeed the gravitomagnetic London moment for Lead is on the average (in- and above-ring location) about 84% of the one obtained with Niobium. This is a very good fit with our Cooper-pair/bulk density ratio prediction.

## 5. CONCLUSIONS

About 250 test runs were performed with liquid helium to investigate the gravitomagnetic London moment for YBCO, Niobium and Lead. As predicted, Niobium gave the strongest accelerations in the tangential direction when it was angularly accelerated. The fact that the above-ring signals (lead and niobium) were about 77-90% of the in-ring values is a strong indication, that we detected an acceleration field emitted from the superconductor. All mean values are about 3.3 times above the noise level. Both Niobium and Lead coupling factors are about a factor of 13 higher than predicted by theory. No effect was seen for YBCO and BSCCO, which is to be expected from our theoretical predictions even considering the offset factor of 13 (YBCO would then produce a coupling constant inside the ring of $-7x10^{-9}$ $s^2$, which is well below our measurement resolution of $\pm 2x10^{-8}$ $s^2$). All measurements were performed by high-precision accelerometers inside an evacuated, mechanically decoupled Faraday cage. All non-transition measurements were performed in a differential setup by subtracting mechanical movements of the sensors induced by the apparatus or from ground floor noise.

Our measurements can be summarized as follows:

– An acceleration field was found to be induced by applying angular accelerations to a superconductor. The field produced is directly proportional to the applied acceleration with a correlation factor higher than 0.96. All mean values are 3.3 times above the facility noise level.

– The gravitational field is emitted from the superconductor and follows the laws of field propagation and induction similar to those of electromagnetism as formulated in linearized general relativity.

- There is a reasonable fit between the field strength observed and predicted by the ratio of Cooper-pair and bulk density for the tested superconductors (Niobium, Lead, YBCO and BSCCO).

- No effect could be seen above the superconductor's critical temperatures and for YBCO and BSCCO throughout all temperature ranges within the accuracy of our experimental setup.

- Gravitational peaks were observed when the superconductor passed its critical temperature while it was rotating. Their sign changed with the orientation of the angular velocity.

Although the signal-to-noise ratio is still low at 3.3, our findings show strong evidence that the gravitomagnetic London moment exists as predicted. If our results are repeated and confirmed, our experiment demonstrates

- For the first time, non-Newtonian gravitational and gravitomagnetic fields of measurable magnitude were observed in a laboratory environment.

- The existence of the gravitational Faraday law was shown.

- The hypothesis that large gravitomagnetic fields may be responsible for the Tate Cooper-pair mass anomaly was found to be supported.

The reported results are very different from previous claims in the literature from Podkletnov claiming gravitational shielding effects above rotating superconductors[21,22]. As we have not observed any change in the vertical sensors (± 5 µg) above any superconductors during their phase transition and during rotation, our results even put new limits on any possible shielding effects (effect must be < 0.0005% compared to claims of up to 2% of weight change for samples above a rotating superconductor).

If our results are confirmed, they would have a profound impact on present gravitational and coherent matter physics. Although the observed effects are small, they open up a new window to investigate the consequences of general relativity and cosmology at low-energies. The results could be used to produce even larger gravitational fields in laboratories.

We hope to have triggered enough interest to stimulate other groups to replicate our results in order to consolidate our findings regarding the experimental existence of the gravitomagnetic London moment and its major consequences.

## 6. ACKNOWLEDGEMENTS

This research was partly sponsored by the European Space Agency under GSP Contract 17890/03/F/KE and by the Air Force Office of Scientific Research, Air Force Material Command, USAF, under grant number FA8655-03-1-3075. The U.S. Government is authorized to reproduce and distribute reprints for Governmental purposes notwithstanding any copyright notation thereon. The team was supported by K. Hense, I. Vasiljevich, B. Seifert and R. Sedmik. Also the efforts from T. Sumrall, M. Fajardo, D. Littrell, I. Wysong and B. Flake to support our experimental efforts are greatly appreciated.

**TABLE I Uncertainty Budgets**

| | |
|---|---|
| Accelerometer noise level at 10 Hz sampling rate – Vertical | ± 5 µg |
| Accelerometer noise level at 10 Hz sampling rate – Radial | ± 15 µg |
| Accelerometer noise level at 10 Hz sampling rate – Tangential | ± 15 µg |
| Keithley Nanovoltmeter (Accuracy converted into g units) | ± 0.45 µg |
| SC velocity measurement | < 1% |
| Time measurement resolution | 1 ms |
| SC temperature | ± 0.5 K |
| Magnetic field measurement resolution at 10 Hz sampling rate | < 0.5 nT rms |
| Induced Gravitational Field / Applied Acceleration Coupling Factor Noise Level | $\pm 2.10^{-8}$ $s^2$ |

**TABLE II Summary of Mean (Upper Limit) Coupling Factors for High and Low $T_c$ Superconductors**

| Material | In-Ring Coupling Factor [$s^2$] | Nr of Samples | Above-Ring Coupling Factor [$s^2$] | Nr of Samples |
|---|---|---|---|---|
| BSCCO (2212) | $< \pm 2.10^{-8}$ | - | $< \pm 2.10^{-8}$ | - |
| YBCO | $< \pm 2.10^{-8}$ | - | $< \pm 2.10^{-8}$ | - |
| Lead | $-8.63 \pm 0.45 \times 10^{-8}$ | 24 | $-6.7 \pm 0.32 \times 10^{-8}$ | 26 |
| Niobium | $-9.64 \pm 0.28 \times 10^{-8}$ | 111 | $-8.71 \pm 0.24 \times 10^{-8}$ | 83 |

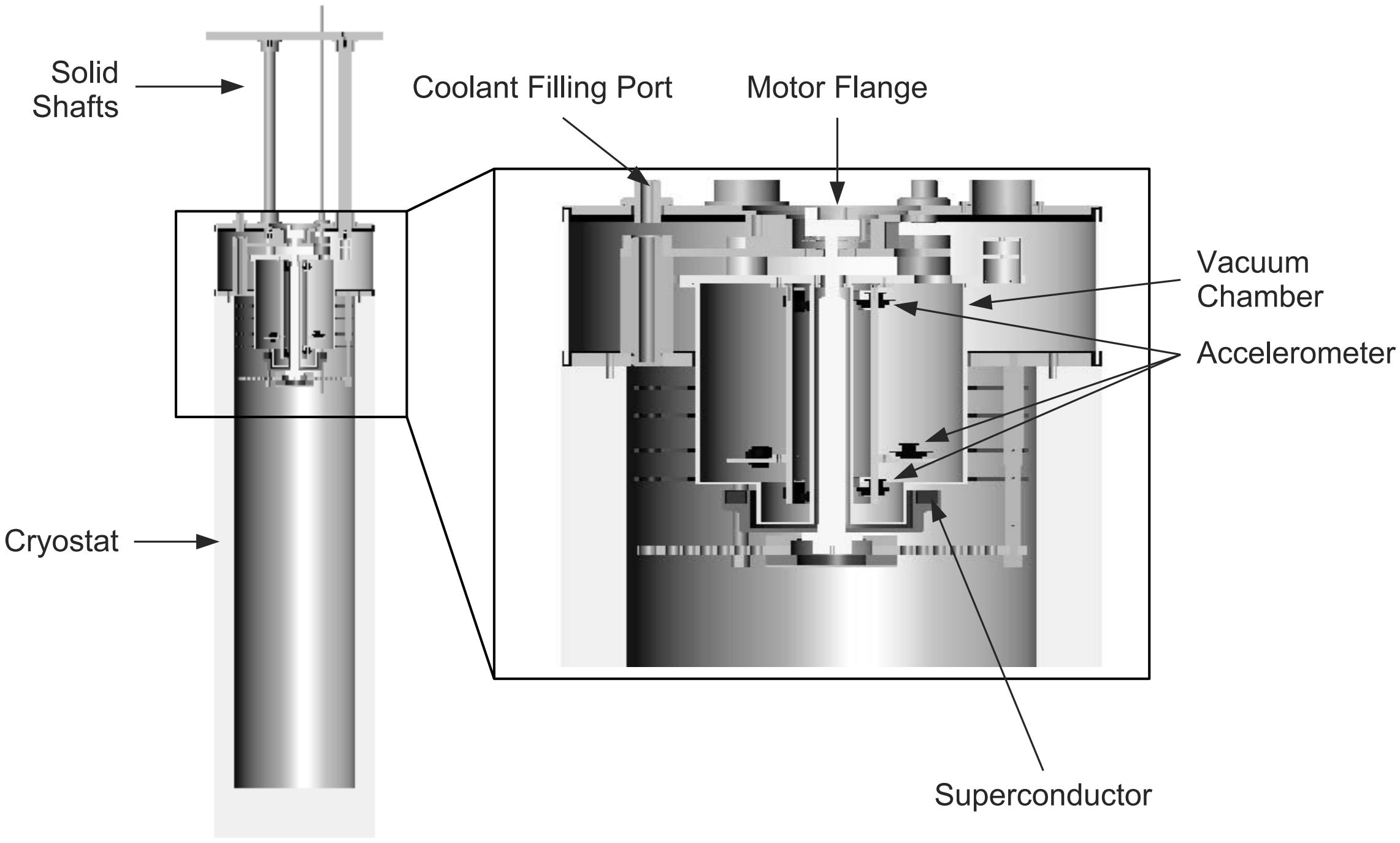

**FIG. 1** Experimental setup

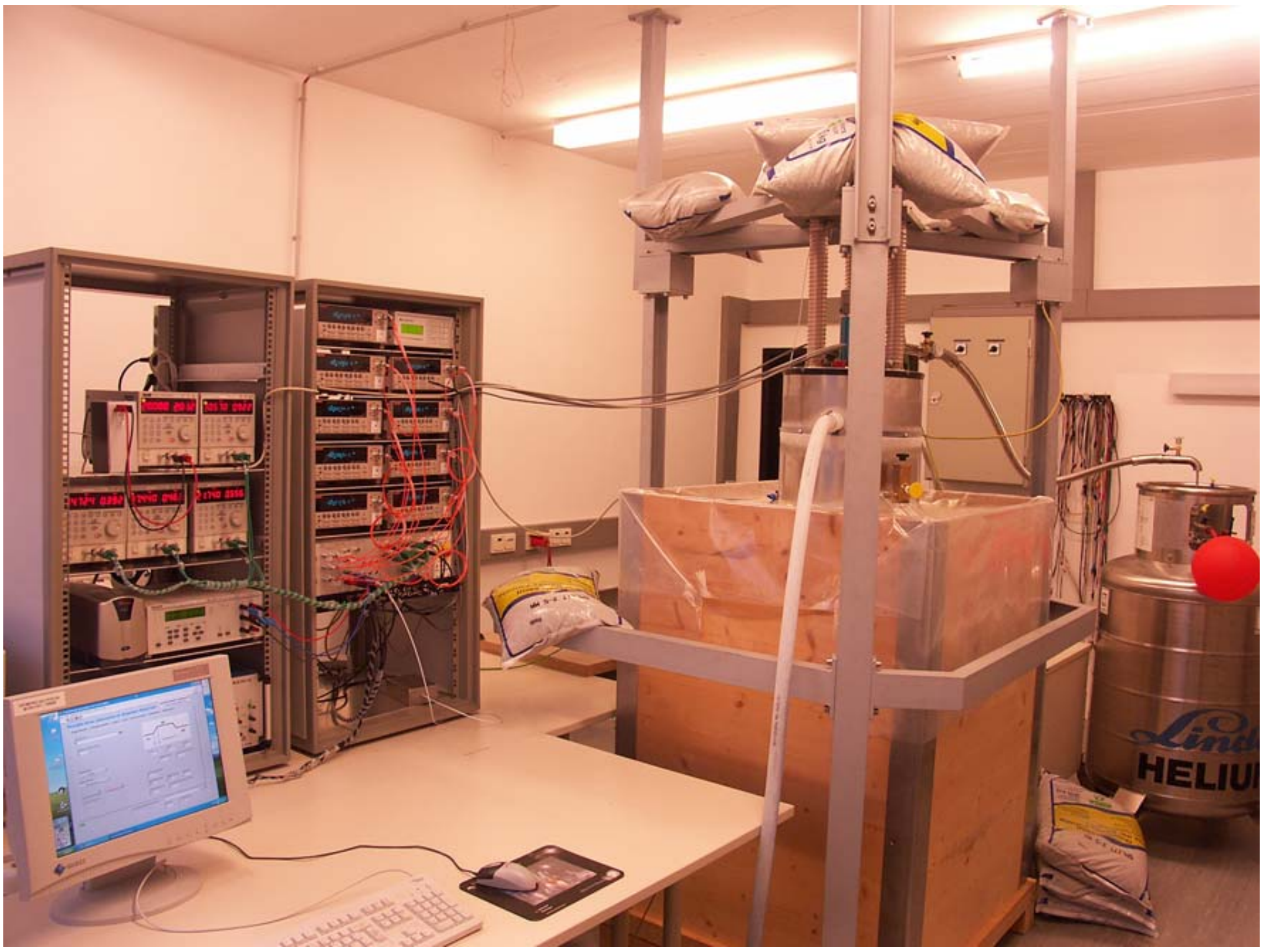

**FIG. 2** Facility during test campaign

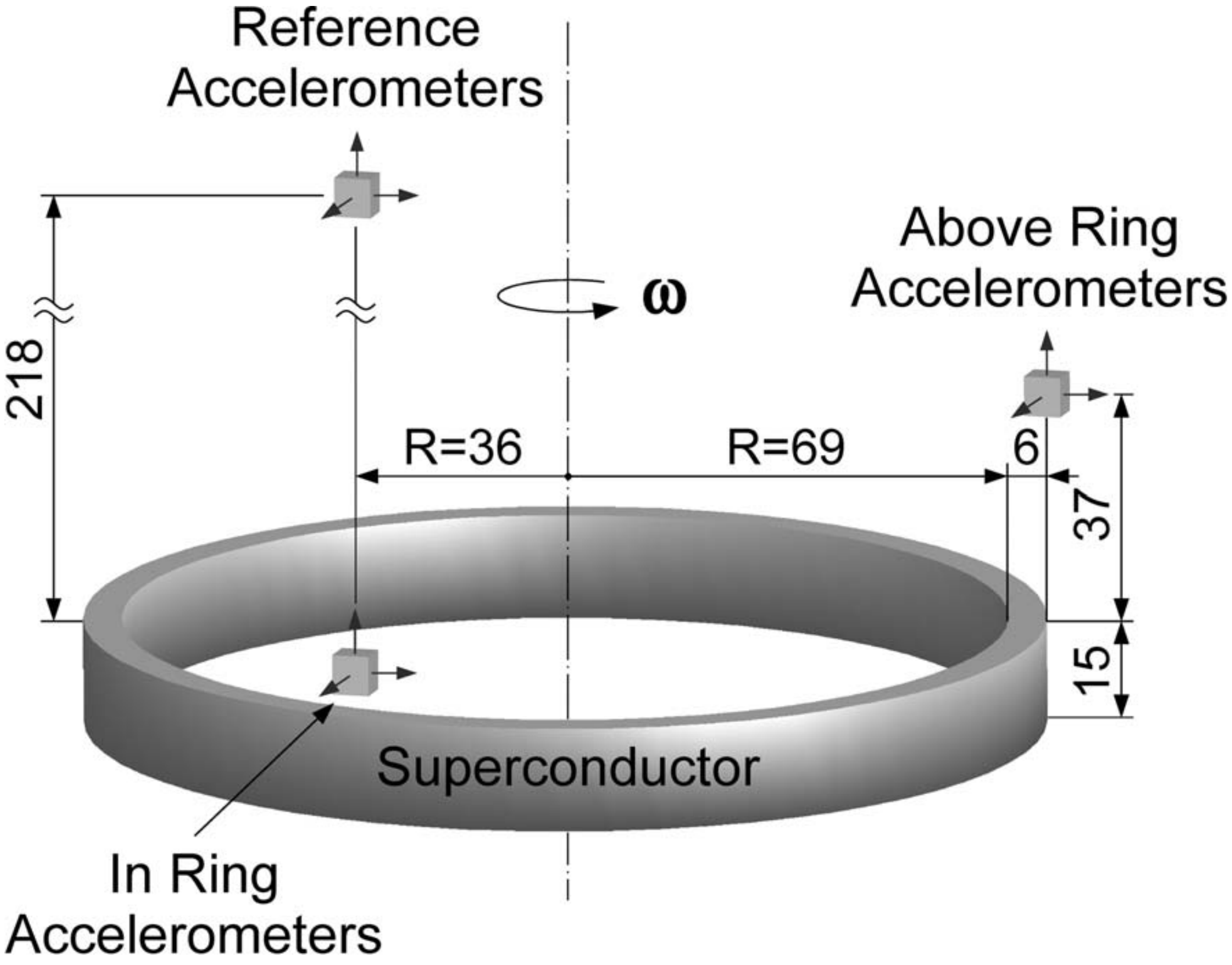

**FIG. 3** Rotating superconductive ring and accelerometer positions (all measures in mm)

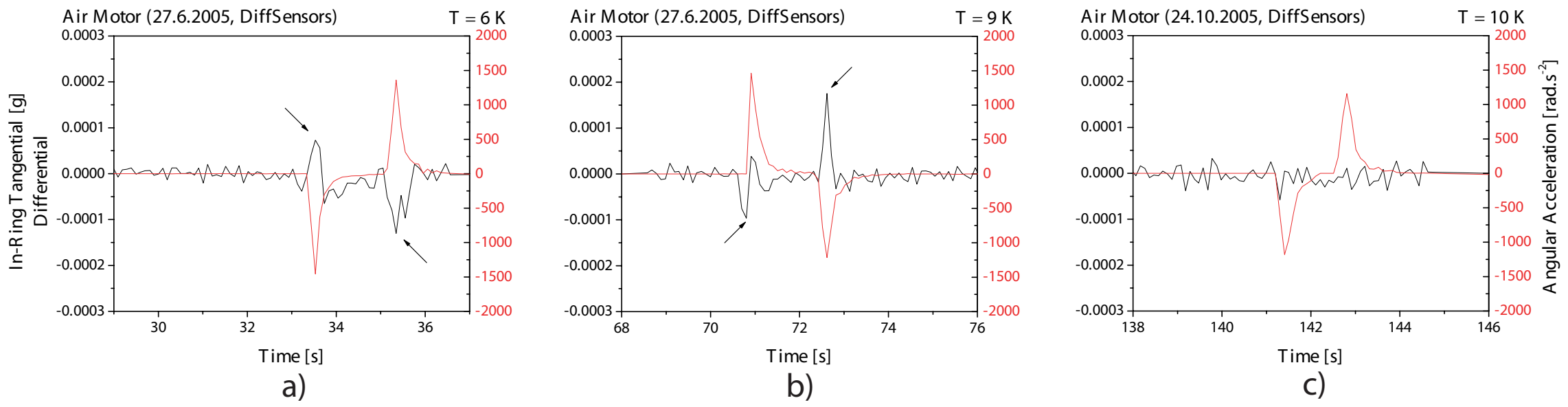

**FIG. 4** Niobium differential in-ring tangential sensor runs at 6, 9 and 10 K (air motor); arrows indicate induced acceleration peaks (acceleration signal for angular velocity > 350 rad.s$^{-1}$ damped by factor 5 to reduce resonance vibrations of motor axis)

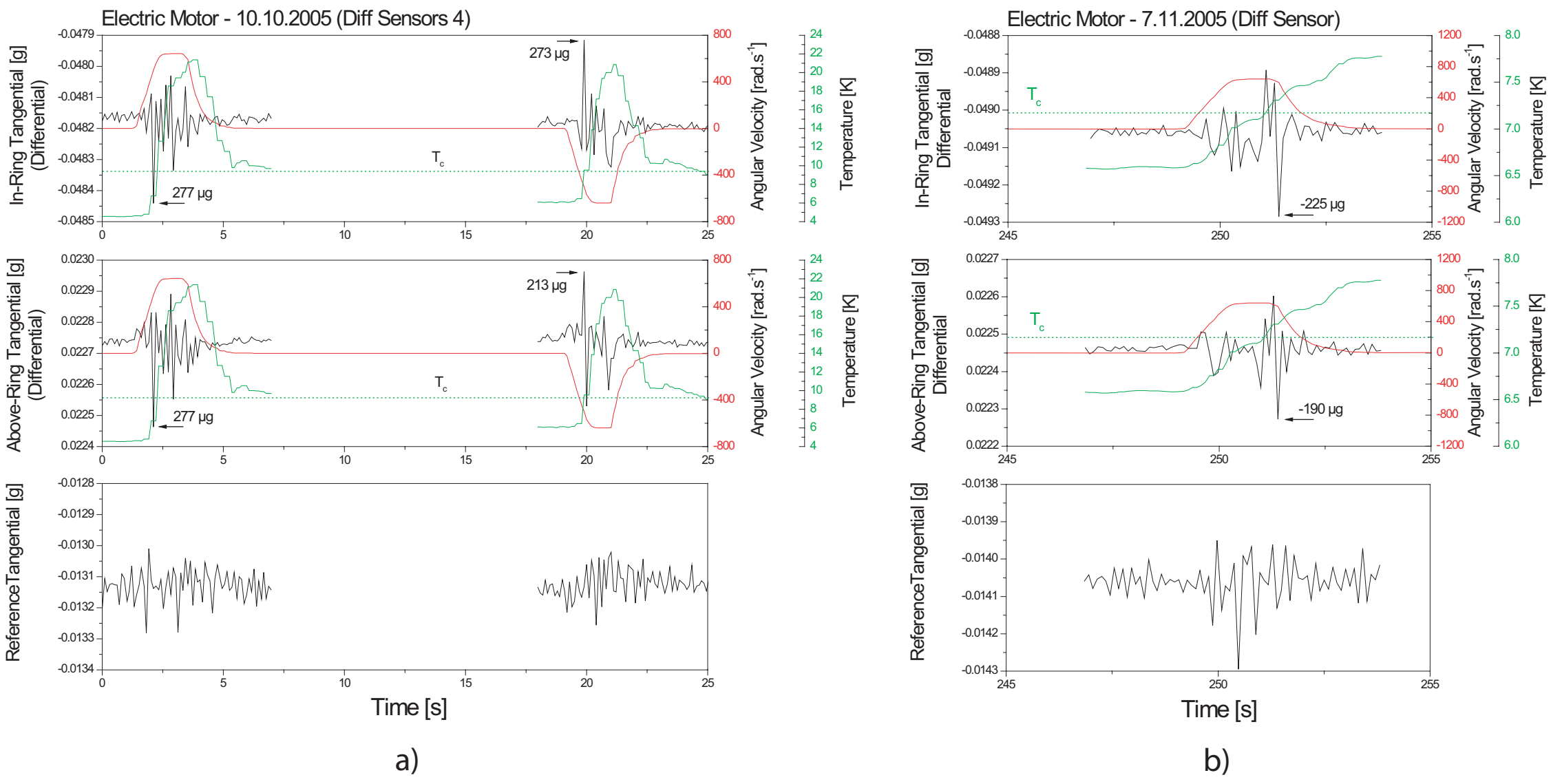

**FIG. 5** All tangential sensor runs during temperature transition for Niobium (left, 2 runs shown) and Lead (right)

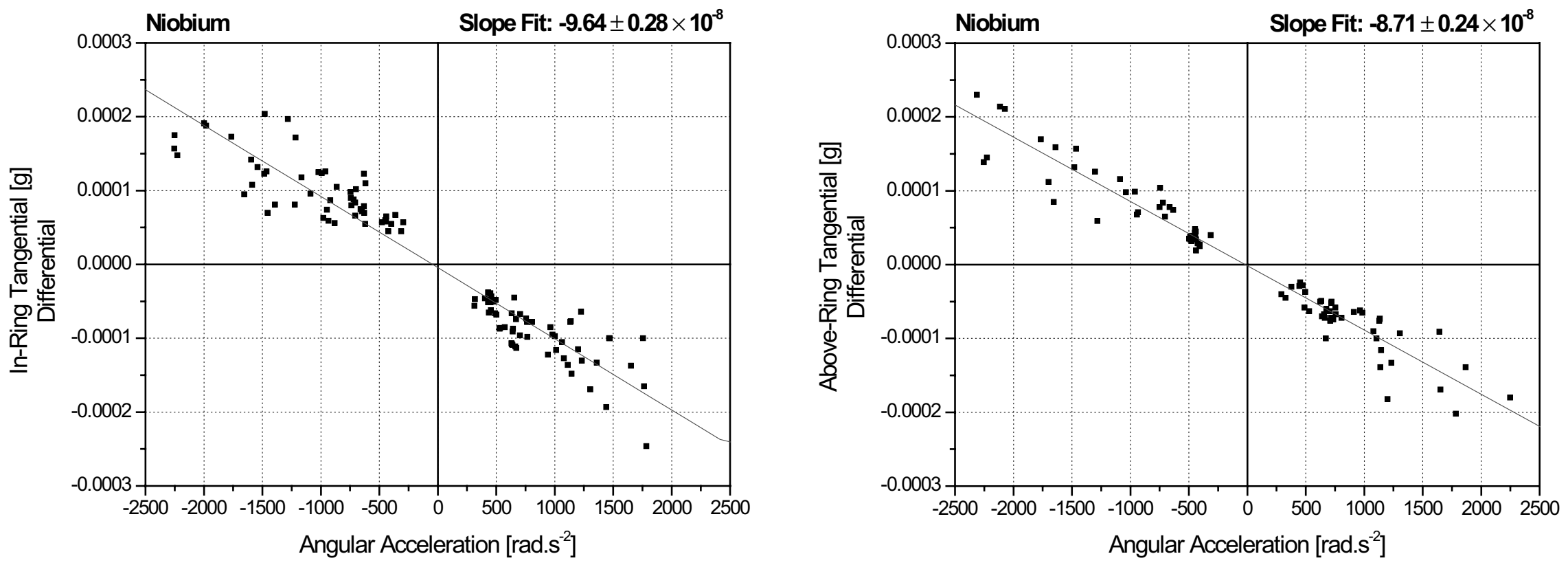

**FIG. 6** Niobium coupling factor slope for in-ring (left, 111 measurements with correlation factor -0.96) and above-ring (right, 83 measurements with correlation factor -0.97) sensor position